\documentclass[fleqn,usenatbib]{mnras}

\usepackage{newtxtext,newtxmath}

\usepackage[T1]{fontenc}

\DeclareRobustCommand{\VAN}[3]{#2}
\let\VANthebibliography\thebibliography
\def\thebibliography{\DeclareRobustCommand{\VAN}[3]{##3}\VANthebibliography}

\usepackage{graphicx}	\usepackage{amsmath}	\usepackage{enumitem}
\usepackage{papermacs}
\usepackage{multicol}
\usepackage{graphicx}
\usepackage{subcaption}
\usepackage{xcolor}

\newcommand{\vphot}{\ensuremath{v_\text{phot}}\xspace}
\newcommand{\vmin}{\ensuremath{v_\text{min}}\xspace}

\newcommand{\nnni}{\ensuremath{\nuc{56}{Ni}}\xspace}

\graphicspath{{./figs/}{./myprogs/}}

\title{Direct Analysis of the Broad-Line SN 2019ein: Connection with
  the Core-Normal SN~2011fe}

\author[Z. Yarbrough et al.]{Zach Yarbrough$^{1,2}$, E. Baron$^{2,3}$,
  James M. DerKacy$^{4,2}$, I.~Washington$^{2}$, P.~Hoeflich$^{5}$,
  Anthony Burrow$^{2}$
  \\
${^1}${Department of Physics \& Astronomy, Louisiana State
  University, Baton Rouge, LA 70803, USA}\\
${^2}${Homer L. Dodge Department of Physics and Astronomy, University of 
Oklahoma, 440 W. Brooks, Rm 100, Norman, OK USA}\\
${^3}${Hamburger Sternwarte, Gojenbergsweg 112, 21029 Hamburg, Germany}\\
${^4}${Department of Physics, Virginia Tech, Blacksburg, VA
  24061, USA}\\
${^5}${Department of Physics, Florida State University, 77 Chieftan Way, 
  Tallahassee, FL 32306, USA}
}

\date{Accepted 2023 March 08. Received 2023 February 22; in original form 2022 August 15}

\pubyear{2023}

\begin{document}
\label{firstpage}
\pagerange{\pageref{firstpage}--\pageref{lastpage}}
\maketitle

\begin{abstract}

Type Ia supernovae (SNe~Ia) are important cosmological probes and
contributors to galactic nucleosynthesis, particularly of the iron
group elements. To improve both their reliability as cosmological
probes and to understand galactic chemical evolution, it is vital to
understand the binary 
progenitor system and explosion mechanism. The classification of SNe~Ia
into Branch groups has led to some understanding of the 
similarities and differences among the varieties of observed
SNe~Ia. Branch groups are defined by the pseudo equivalent widths of the two
prominent \ion{Si}{II} lines, leading to  four distinct groups:
Core-Normal (CN), Shallow-Silicon (SS), Cool (CL), and Broad-Line
(BL).   However, partly due to small sample size, little work 
has been done on the BL group. We
perform direct spectral analysis on the pre-maximum spectra of the
BL SN~2019ein, 
comparing and contrasting to the CN SN~2011fe. Both
SN~2019ein and SN~2011fe were first observed spectroscopically within
two days of discovery, allowing us to follow the
spectroscopic evolution of both supernovae in detail. We find that the
optical depths of the primary features of both the CN and BL
supernovae are very similar, except that there is a Doppler shift
between them. We further examine the BL group  and show
that for nine objects with pre-maximum spectra in the range $-6$ --- $-2$
days with respect to $B$-maximum 
all the emission peaks of the \ion{Si}{II} $\lambda 6355$
line of BL are blueshifted pre-maximum, suggesting a possible
classification criterion. 
\end{abstract}


\section{Introduction}
\label{sec:intro}

While it is generally agreed that Type Ia supernovae (SNe Ia) are
explosions of a carbon/oxygen 
(C/O) white dwarf in a binary system \citep{Hoyle1960}, the 
full nature of the progenitor system, especially the nature of the
secondary star, is still unclear \citep[for a review
  see][]{Maoz:2014}. Theoretical scenarios for the nature of the
explosion were
originally  divided into the single-degenerate (SD) and
double-degenerate (DD) scenarios. In the SD
scenario, the companion is either a main-sequence star or an evolved,
non-degenerate companion like a red giant or He-star
\citep{Iben:Tutukov:1984}. Through accretion, the white dwarf
approaches the Chandrasekhar mass, and eventually explodes via a
deflagration-to-detonation transition
\citep{Khokhlov:1991,Hoeflich1995,Hoeflich:2002,Hoeflich:2006:NP-Review}. In
the DD scenario, the companion is also a white
dwarf, where the combined mass of the system equals or exceeds the
Chandrasekhar mass and the explosion is triggered by the merger of the
two WDs \citep{Iben:Tutukov:1984,Webbink1984}. 

More recently the progenitor scenarios have been separated into two
categories. Near-Chandrasekhar mass scenarios envision a white dwarf
accreting material from a companion, or the 
merger of a WD with the core of an evolved star 
\citep[the core-degenerate
scenario][]{Kashi2011,Soker2014}. In the sub-Chandrasekhar mass
double detonation scenario
the central sub-Chandrasekhar mass WD is detonated due to compression
from the detonation of a low mass helium shell on its 
surface
\citep{Woosley:1994:Weaver:HeDet,Livne:1995:Arnett:HeDet,Shen2018,Polin2019}. 
The
  sub-Chandrasekhar helium detonation has been attributed to
  SN~2019ein \cite{Xi:2022:19ein}, SN~2020jgb
  \citep{Liu_etal_2022_20jgb}, and SN~2016dsg
  \citep{Dong_etal_2022_16dsg}.  The near
  Chandrasekhar mass scenario has received a boost from recent JWST
  observations of SN~2021aefx
  \citep{Derkacy_etal_2023_01aefx,Kwok_etal_2023}.

SNe~Ia  are important cosmological
probes due to the 
fact that their luminosity is empirically related to the shape of the
light curve \citep{Phillips1993,Phillips:1999}. Since the relationship
to correct the peak brightness to the lightcurve shape  (the Phillips
relation) is purely
empirical, it is important to be able to identify any systematic
biases that may be associated with variations in progenitors and
explosion mechanism. Determining the  nature of the progenitor system
of SNe~Ia is 
a crucial piece of this puzzle and while the exact progenitor
system(s) remaining unknown, understanding the empirical relations that
link different SNe~Ia can help shed light on the progenitor system.

\citet{Branch2006} defined a set of four spectroscopically defined
groups based upon where the locus of supernovae fall in a plot of the
pseudo equivalent width of \ion{Si}{II} $\lambda 5972$ versus
\ion{Si}{II} $\lambda 6355$. 
They established four distinct groups:
Core-Normal (CN), Broad-Lines (BL), Shallow-Silicons (SS), and Cools
(CL). The group assignments were somewhat arbitrary, but were guided
by where the SN fell in the width luminosity relation
\citep{Phillips1993,Phillips:1999}, with CN and BL having values of
$\Delta m_{15}(B) \sim 1.1$, where $\Delta m_{15}(B)$ is the number of
$B$ magnitudes that the light curve declines by in the 15~days immediately 
after B-maximum
\citep{Phillips1993}. SS were associated with slow declining, more
luminous SNe and CL were associated with fast slow declining, dimmer
SNe. The Branch diagram was further explored in
\citet{Blondin_etal_2012}, who showed that there existed a strong
correlation between the ratio of the $\lambda 5972$ and $\lambda 6355$
pseudo-equivalent widths. 
\citet{Burrow2020}, using the full Carnegie Supernova Project
I+II sample
\citep{Folatelli_etal_2013,Krisciunas:2017,Phillips:2019,Hsiao:2019},
showed that these groups are statistically robust and produced
a model to probabilistically determine Branch group membership.
They also showed that the pEW of
$\lambda 5972$ is well correlated with the luminosity width parameter
$s_{BV}$ defined by \citet{Burns_etal_2014}.
Currently, there is no strong correlation between the intrinsic
luminosity of the SN and its explosion mechanism, although
\citet{Polin2019} identified fast decliners with sub-Chandrasekhar
helium detonations.
The BL group, including
class defining events such as SNe~1984A and 2002bo has not been
well studied.

Other SNe~Ia spectral classification schemes have subdivided SNe~Ia
on different criteria than the Branch scheme. 
\citet{Benetti:2005} broke SNe~Ia into three groups, based primarily
on the change in \ion{Si}{II} $\lambda 6355$ velocity after maximum light
(measured in \kmps per day):
High Velocity Gradient (HVG), including SN 1984A and SN~2002bo;
Low Velocity Gradient; and a Faint group, suggesting that there was a
physical difference among them. \citet{Wang:2009} broke SNe~Ia into
two groups, based on whether their \ion{Si}{II} $\lambda 6355$ velocities
were greater than or less than 11,800~\kmps at maximum light:
High-Velocity and Normal and obtained better distance estimates by
correcting for membership in each class. SN~1984A and SN~2002bo
fell into the High-Velocity group. \citet{Burrow2020} show that
characteristic velocities for the \ion{Si}{II} $\lambda 6355$ line at
maximum light fall in the range 9,000 -- 12,000~\kmps and that BL have
velocities in the range 12,000 -- 16,000~\kmps although some CN also
have velocities approaching 15,000~\kmps.

SN 2019ein was discovered on 2019 May 1.5 (UT) in NGC~5353
by the Asteroid Terrestrial-impact Last Alert System (ATLAS) project
in the cyan-ATLAS band at 18.194~mag \citep{Tonry:ATLAS19ieo:2019}.
The first spectrum was obtained on May 2.3 (UT) by the Las Cumbres
Observatory Global SN Project (GSP), showing SN~2019ein to be
an 02bo-like or BL\footnote{While the 02bo-like and BL classification
are not identical, henceforth we will identify SN~2019ein by its
Branch group.} SN Ia at about two weeks before maximum light
\citep{Burke:2019ein:2019}.

\citet{Pellegrino:2020:19ein} presented GSP observations that revealed
\ion{Si}{II} extending out to greater than 25,000~\kmps as measured
from the \ion{Si}{II} $\lambda 6355$ line, and that the emission peaks
of the P-Cygni profiles were blue-shifted by up to 10,000~\kmps.  A
detailed study of photometric and spectroscopic data of SN~2019ein
by \citet{Kawabata:Maeda:2019ein:2020} concluded that the outermost
layer of the progenitor is an O-Ne-C burning layer extending
to 25,000 -- 30,000~\kmps.
\citet{Patra:Yang:2019ein:2022} performed spectropolarimetry on
SN~2019ein ruling out global asphericity of the ejecta.
\citet{Xi:2022:19ein} present spectra and photometry
of SN~2019ein and find that the SN was likely produced in a
sub-Chandrasekhar explosion owing to the inferred low luminosity and
nickel mass, combined with the high ejecta velocities.

\subsection{Motivation}
\label{sec:moto}

We perform a direct spectroscopic analysis of the pre-maximum spectra
of SN~2019ein and compare our findings with the direct
analysis of the pre-maximum spectra of SN~2011fe.
SN~2011fe is a well-observed, well-studied core-normal supernova, that was caught 11
hours after explosion
\citep{nugent:2011:11fe,bloom:2012:11fe,brown:2012:11fe,dessart:2014:11fe,Pereira:2013:11fe,hsiao:2013:11fe,li:2011:11fe,mazzali:2014:11fe,Baron:2015,zhang:2016:11fe,derkacy:2020:11fe}.
Since SN~2019ein is a
BL supernova \citep{Pellegrino:2020:19ein} and SN~2011fe is a 
CN supernova \citep{Parrent:2012:11fe}, the comparisons are helpful for our understanding of
the similarities and differences that are captured in the Branch group
classification scheme \citep{Branch2006,Burrow2020}. Both supernovae were
discovered very early and thus we compare spectra obtained at days
(-14, -10, -6, -4, 0) for SN~2019ein and at days (-13, -10, -7, -3, 0)
for SN~2011fe, where day 0 is the time of maximum in the $B$ band.

We aim to compare CN and BL spectra at both similar and  
different epochs in their evolution relative to each other.
These comparisons should provide insight into differences into 
the progenitor/explosion mechanisms, although we do attempt to 
associate each group with a specific system. Instead, we seek
to describe the observed features that are distinct to the BL sub-class. This
behavior will then need to be reproduced by theoretical models.

\section{Methods}
\label{sec:meths}

\begin{figure*}
\centering
\subcaptionbox{SN~2019ein}{\includegraphics[width=0.45\linewidth]{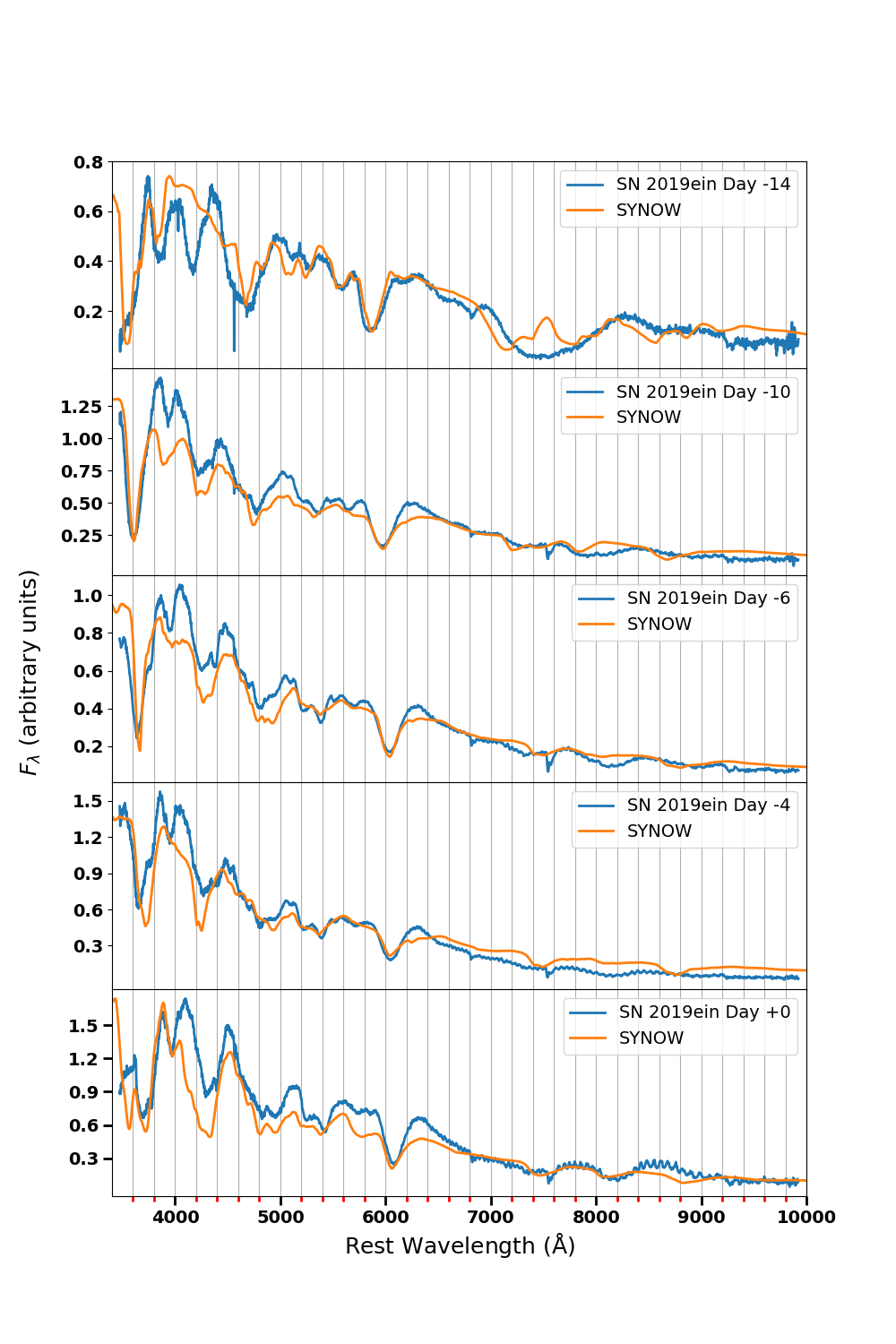}}
\subcaptionbox{SN~2011fe}{\includegraphics[width=0.45\linewidth]{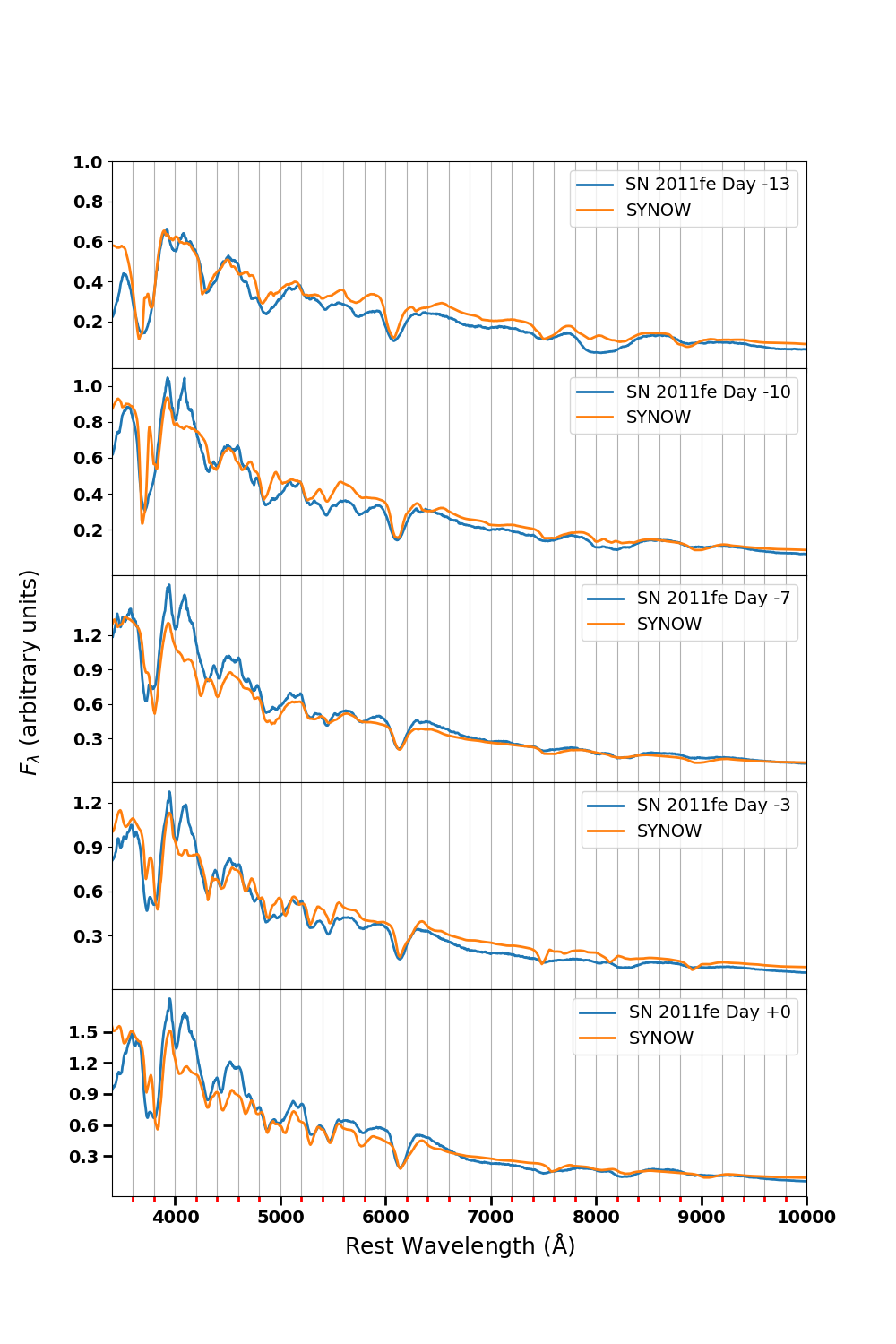}}
\caption{SYNOW fits to pre-maximum spectra of SNe~2019ein and
    2011fe.}
\label{fig:fits}
\end{figure*}

We
use the spectral modeling tool SYNOW \citep{Fisher:2000:PhD} to
generate fits to SN~2019ein (see \autoref{fig:fits}) at the five
pre-maximum epochs.
SYNOW
assumes that the optical depth of a specified reference line of a
given ion follows an exponential decay \[ \tau(v) = \tau_0
e^{-({v-\text{max}(v_\text{ph},v_\text{min}))/v_e}}, \] where
  $v_\text{ph}$ is the velocity of the photosphere, $v_\text{min}$ is
  the specified minimum velocity for this component of the ion's
  lines, and $v_e$ is a parameter specifying the rate at which the
  optical depth falls off with velocity. $v_\text{ph}$ is a global
  fitting parameter, whereas $v_\text{min}$ is specified for each
  feature. In the case that  $v_\text{min} > v_\text{ph}$ the feature
  is said to be detached. The atmosphere is assumed to obey the
  Schuster-Schwarzschild approximation \citep{mihalas78sa}, to be totally opaque below
  $v_\text{ph}$, and the source function is taken to be that of
  coherent scattering in the Sobolev approximation
  \citep{sob60}. SYNOW includes the effects of multiple scattering and
  the relative line strengths are given by the Boltzmann factor with
  an ion specific parameter $T_e$. A
  particular ion may be represented in a fit by more than one
  feature and we focus here on observed features that require fits
  with both photospheric and detached components.

After 
fitting SN~2019ein, we used the velocity information in
\citet{Parrent:2012:11fe} as a starting point to make corresponding
fits to the pre-maximum spectra of SN~2011fe. That is, we extracted the
velocities obtained by \citet{Parrent:2012:11fe} and used them to
produce preliminary SYNOW fits, and then adjusted those fits until we
were satisfied with their quality.    \autoref{fig:fits}
shows the SYNOW fits to both supernovae. Our goal was to generally
reproduce the velocity extents of the prominent spectral features
(\ion{Si}{II}, \ion{S}{II}, \ion{Ca}{II}, \ion{Fe}{II}, \ion{O}{I}, \ion{C}{II}, and \ion{Mg}{II}), in
order to compare and contrast the two supernovae using the parameters
obtained by SYNOW.

\section{Discussion}
\label{sec:disc}

\subsection{Interpretation of Results}
\label{sec:interp}

\begin{figure*}
\centering
\includegraphics[scale=0.5]{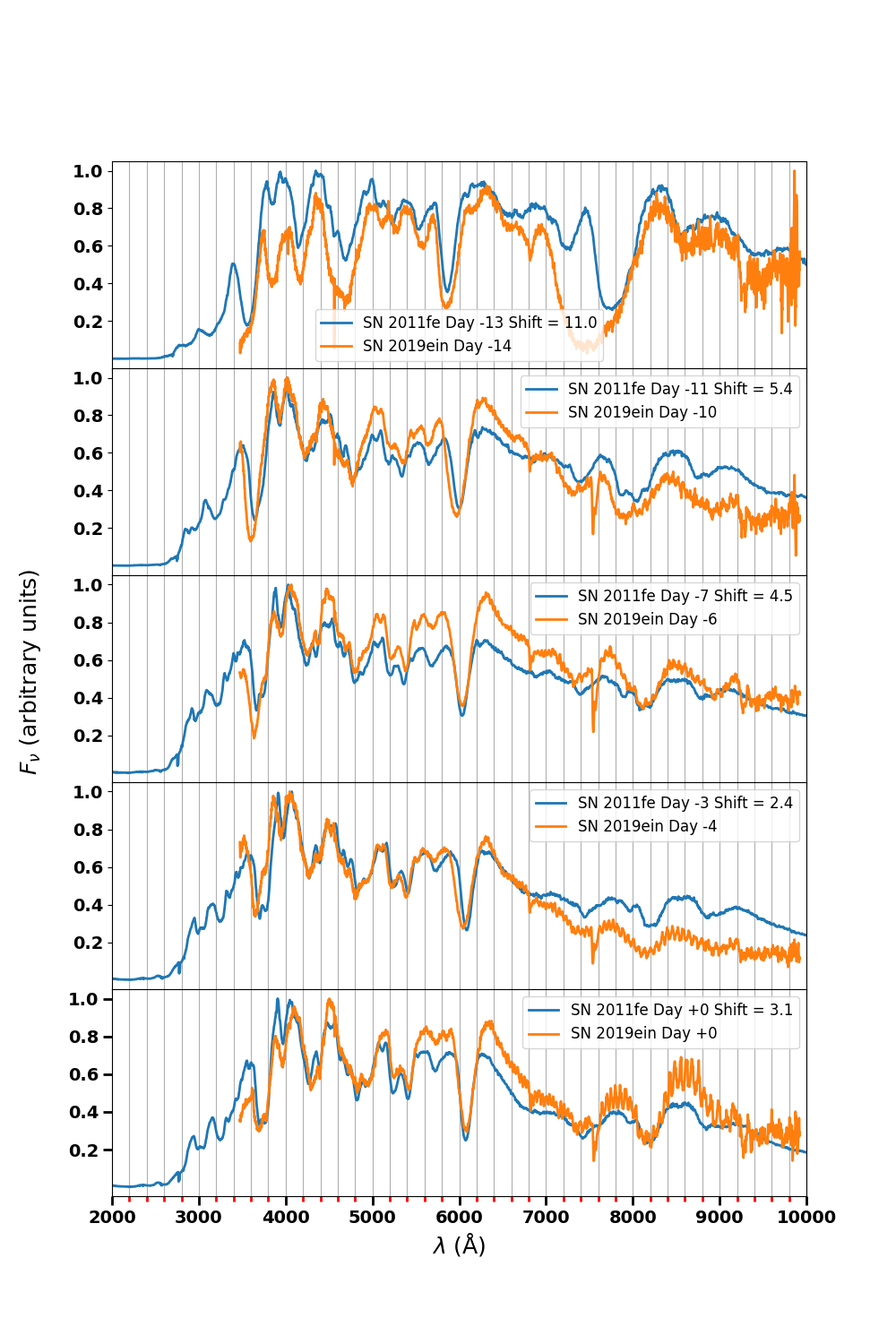}
  \caption{Comparison of of the observed SN~2019ein corrected to the
    rest wavelength with the observed spectra of SN~2011fe, where the SN~2011fe
    spectra have been blue shifted by the amounts indicated in the legend
    in units of $1000$~\kmps.}
  \label{fig:redshifts}
\end{figure*}

\autoref{fig:redshifts}
shows that simply blue-shifting the observed pre-maximum spectra of SN~2011fe
to match the pre-maximum spectra of SN~2019ein 
shows remarkable alignment for almost
all features.
This similarity of the spectra is
striking and indicates that SN~2019ein, up until maximum light, is
physically (at least in terms of the total optical depth  of the ions that
produce the main spectral features),  very
similar to SN~2011fe, with the only difference being the shifted velocities. 

\begin{figure}
\centering
  \includegraphics[scale=0.5]{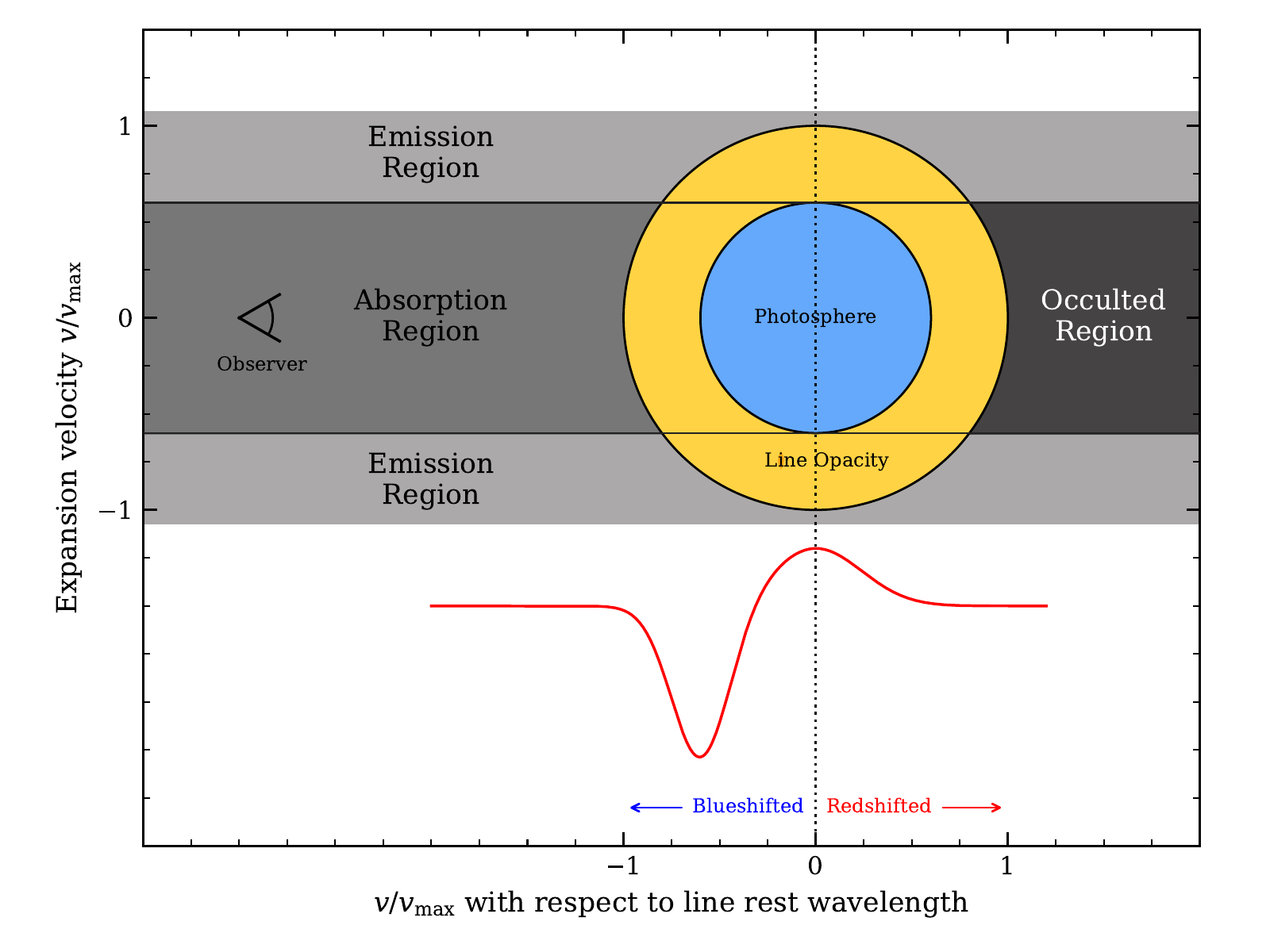}
  \caption{A cartoon illustrating the formation of P-Cygni line
    profiles. Adapted from \citet{Czekala_2011}.}
  \label{fig:pcy-cartoon}
\end{figure}

The formation of P-Cygni lines has been discussed for decades
\citep[see for example,][and references
  therein]{Lucy_1971,Surdej_1979,Castor_Lamers_1979,Wagenblast_etal_1983,Jeffery_Branch_1990,Kasen_etal_2002}. \autoref{fig:pcy-cartoon}
illustrates the basic mechanism of P-Cygni line profile formation and
clearly shows that the emission peak should be at the rest wavelength
of the line for an isolated line that has opacity throughout the line
forming region. 

Yet, \citet{Pellegrino:2020:19ein} noted that in SN~2019ein
the emission peaks were
shifted by up to $10,000$~\kmps and demonstrated that a detached \ion{Si}{II}
component (offset from the photosphere by $2500$~\kmps) could reproduce the \ion{Si}{II}
5970 and \ion{Si}{II} 6355 features at day~$-10$. 
Therefore, since both the absorption and emission peaks of \ion{Si}{II} $\lambda
6355$ are shifted in the observed SN~2019ein spectra, this
  indicates that the community's understanding of how P-Cygni
lines form, as 
  shown in \autoref{fig:pcy-cartoon}, may be incomplete and requires further investigation.

Blue-shifted emission features observed in Type II supernovae have
been suggested to arise from steep density profiles
\citep{Dessart:2005,Dessart:2011, Anderson:2014}. However, in SNe~II
the Balmer lines are so strongly NLTE that SYNOW does a poor job of
modeling them. \citet{Dessart:2005} and \citet{Blondin:2006} argue
that the blueshifted P-Cygni emission is due to an optically thick
continuum near the photosphere. Since SYNOW, a pure line-scattering code
neglects continuum effects, we can not address this.
Nevertheless, as shown below, we obtain very good fits to 
the observed line profiles, so we can infer the velocity extent of the
ions that produce the observed features.

\begin{figure*}
\centering
\subcaptionbox{}{\includegraphics[width=0.45\linewidth]{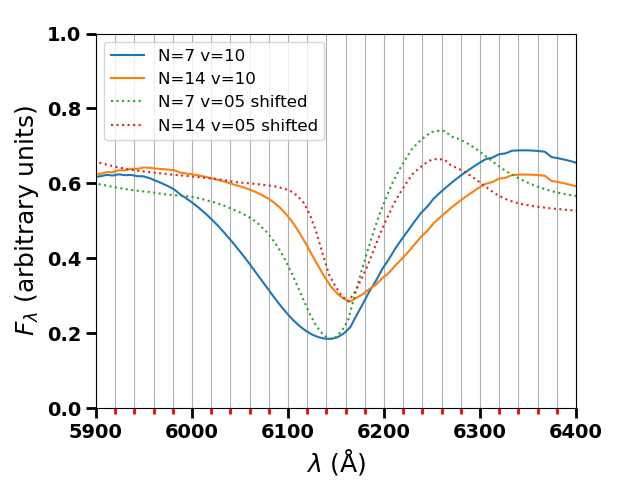}}
\subcaptionbox{}{\includegraphics[width=0.45\linewidth]{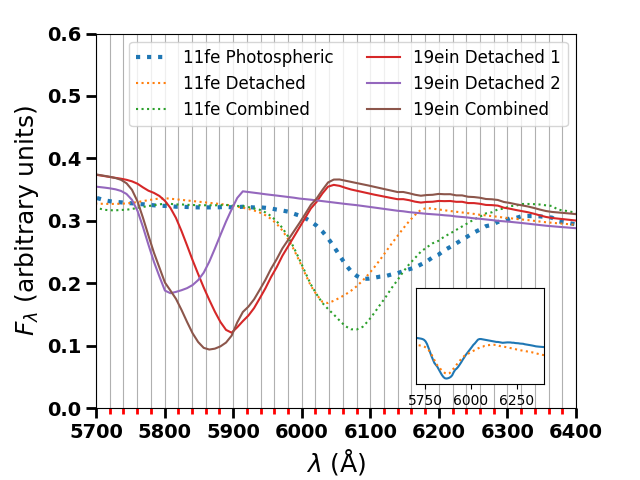}}
  \caption{Panel (a) shows a P-Cygni line
    with $\vphot=10,000$~\kmps  compared to a P-Cygni line
    with $\vphot = 5,000$~\kmps that has been blue shifted by
    $5000$~\kmps. Two density profiles are chosen: $\rho \propto
    v^{-7}$ and $\rho \propto v^{-14}$. Clearly a simple photospheric
    P-Cygni profile 
    does not shift in order to match the observations of SNe~2011fe
    and 2019ein.     Panel (b) shows that detached P-Cygni lines can 
    mimic the observed spectra. The photospheric velocity,
    $v_\text{ph}$, for the 11fe
    fit is 14,000~\kmps, the maximum velocity, $v_\text{max}$ is
    40,000~\kmps, the detatched velocity, $v_\text{detach}$  is
    17,000~\kmps. For 19ein $v_\text{ph} = 18,000$~\kmps,
    $v_\text{max} = 50,000$~\kmps,
    $v_{\text{detach}{_1}}= 25,000$~\kmps,
    and $v_{\text{detach}{_2}} = 30,000$~\kmps. The inset shows the comparison of the
    total combined SYNOW 19ein feature (solid blue line)  with the total combined 11fe
    SYNOW feature blueshifted by $10,500$~\kmps (dotted red line).}
  \label{fig:pcyg}
\end{figure*}

Turning our attention to single P-Cygni profile formation, \autoref{fig:pcyg}(a) shows that a
high velocity photospheric P-Cygni profile can not simply be mimicked
by just Doppler shifting a lower velocity photospheric P-Cygni profile.
Because the prominent emission components are also blue-shifted, when 
the full line profile is Doppler shifted, the emission component of the 
shifted low-velocity now appears
in the absorption trough of the high velocity line. While in the true
SYNOW formulation absorptions trump emissions, \emph{it is important to
understand that even though the observations can be made to match
nearly perfectly by a Doppler shift, an individual P-Cygni feature can not.}

A simple density variation does not remedy this problem either (see again
\autoref{fig:pcyg}(a)). We show a moderate density profile, $\rho \propto v^{-7}$
compared to a moderately steep density profile, $\rho \propto v^{-14}$.
The density variation mostly affects the slope of the
line shape that connects the blue-shift P-Cygni absorption trough to
the P-Cygni emission peak. In both cases (whether there is a
  moderate or steep 
  density gradient), the shifted emission peak is
clearly visible 
in the absorption trough of the higher velocity P-Cygni feature,
which is not seen in the observations. In fact,
\autoref{fig:redshifts} shows that the
shifted SN~2011fe spectrum matches up with that of SN~2019ein quite
well.  Thus, the line formation in SN~2019ein and SN~2011fe must be
more complex than just a simple Doppler shift of
photospheric P-Cygni lines, since changing the density gradient
  in the SYNOW framework does not affect the position of the emission
  peak --- that is, the emission peak is formed at zero velocity (see
  \autoref{fig:pcy-cartoon}).

\begin{figure*}
\centering
\begin{multicols}{2}
\subcaptionbox{}{\includegraphics[width=\linewidth]{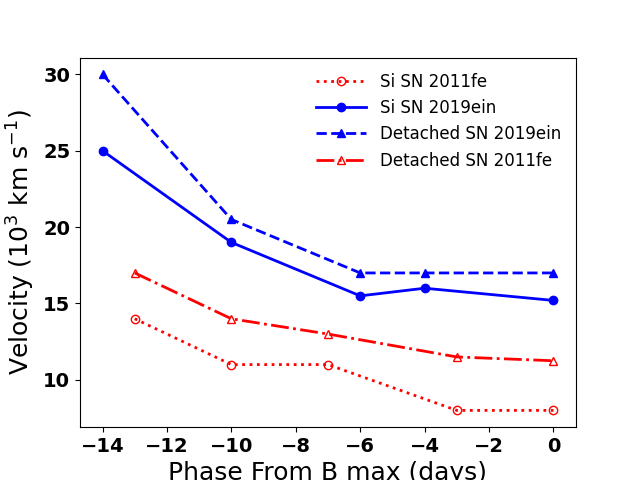}}\par 
\subcaptionbox{}{\includegraphics[width=\linewidth]{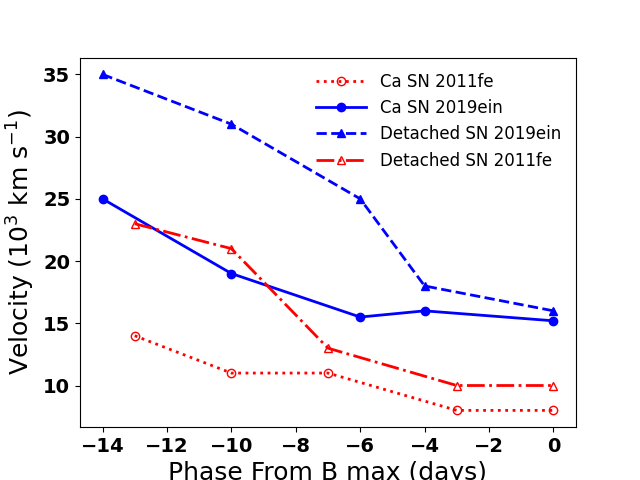}}\par 
\end{multicols}
\caption{SYNOW velocities of \ion{Si}{II} (left) and \ion{Ca}{II}
  (right) obtained for
    SN~2019ein compared to those of SN~2011fe.}
  \label{fig:syn-values:SiCa}
\end{figure*}

\autoref{fig:pcyg}(b) shows the
components of the SYNOW fits of the \ion{Si}{II} $\lambda 6355$ line at the
earliest epoch for both supernovae. In the case of SN~2011fe the
photospheric velocity is 
$\vphot = 14,000$~\kmps. There is a photospheric component to the
\ion{Si}{II} feature and a detached component, $\vmin = 17,000$~\kmps, with
equal values of $\tau = 4$. For SN~2019ein the photospheric velocity is
$\vphot = 18,000$~\kmps. There are two detached components to the
\ion{Si}{II} feature: $\vmin = 25,000$~\kmps ($\tau = 32$) and
$\vmin = 30,000$~\kmps ($\tau = 5$). By maximum light, for SN~2011fe, \vphot 
has dropped to $8,000$~\kmps and the detached feature has 
$\vmin =11,250$~\kmps. Again both components are of equal strength,
$\tau = 2$. 
For SN~2019ein, \vphot
has dropped to $15,200$~\kmps; there is now a photospheric component
and a detached feature with 
$\vmin =17,000$~\kmps. Again both components are of equal strength,
$\tau =8$. Thus, even at maximum light SN~2019ein is significantly
faster in the \ion{Si}{II} photosphere than SN~2011fe.
\autoref{fig:syn-values:SiCa} shows the values obtained for the
photospheric and detached components of the \ion{Si}{II} and
\ion{Ca}{II}. The general trend is clear: the photospheric and
detached components of SN~2019ein are almost universally faster than
those of SN~2011fe. Nevertheless, when shifted, the features are
extremely similar. This suggests that the burning in both supernovae
produces similar amounts of these elements even though the density
structure is quite different.

\subsection{Comparison to Explosion Models}
\label{sec:mods}

\citet{Blondin:2015} modeled the light curve and spectral evolution of
SN~2002bo, using delayed-detonation model DDC15 from
\citet{Blondin:2013}. The transition to a detonation was triggered at
$\rho = 1.8\times 10^7$~\gcm and the central density of the WD was $\rho_c = 2.6\times 10^9$~\gcm. The model does 
a good job of reproducing the light curves and spectra of
SN~2002bo. However, \citet{Pellegrino:2020:19ein} showed that while
DDC15 does a good job of fitting SN~2019ein spectra after about day
-10, the whole P-Cygni structure of the features of SN~2019ein is
blueshifted with respect to DDC15 at day -14, just the effect we focus
on here.

\citet{Lentz:2001} modeled SN~1984A using the delayed detonation models DD21c of
\citet{Hoeflich:1998} and CS15DD3 of \citep{Iwamoto:1999}. Both models
do a good job of reproducing the spectra of SN~1984A. Similarly
\citet{Baron:2015} modeled the evolution of the spectra of SN~2011fe
using the delayed detonation model Z23 of \citet{Hoeflich:2002}, the
transition density was triggered at $\rho = 2.3\times 10^7$~\gcm and
the central density was $\rho_c = 2.0\times 10^9$~\gcm.

\citet{Baron:2012:01ay} model the BL SN~2001ay in detail using the
pulsating delayed detonation model PDD\_11b. The pulsating delayed
detonation mechanism occurs when the deflagration is quenched, prior
to the white dwarf becoming unbound, during the contraction phase a
detonation can occur \citep{Khokhlov_Mueller_Hoeflich_1993}. Besides being a clear BL,
SN~2001ay is an extremely slow decliner, whose brightness is
less than that which would be predicted by the Phillips relation.
Nevertheless, the synthetic spectra do a very good job of reproducing
the observed \ion{Si}{II} line profiles, particularly with respect to
the positions 
of the minimum velocity and the emission maximum.

\begin{figure*}
\centering
\subcaptionbox{}{\includegraphics[width=0.45\linewidth]{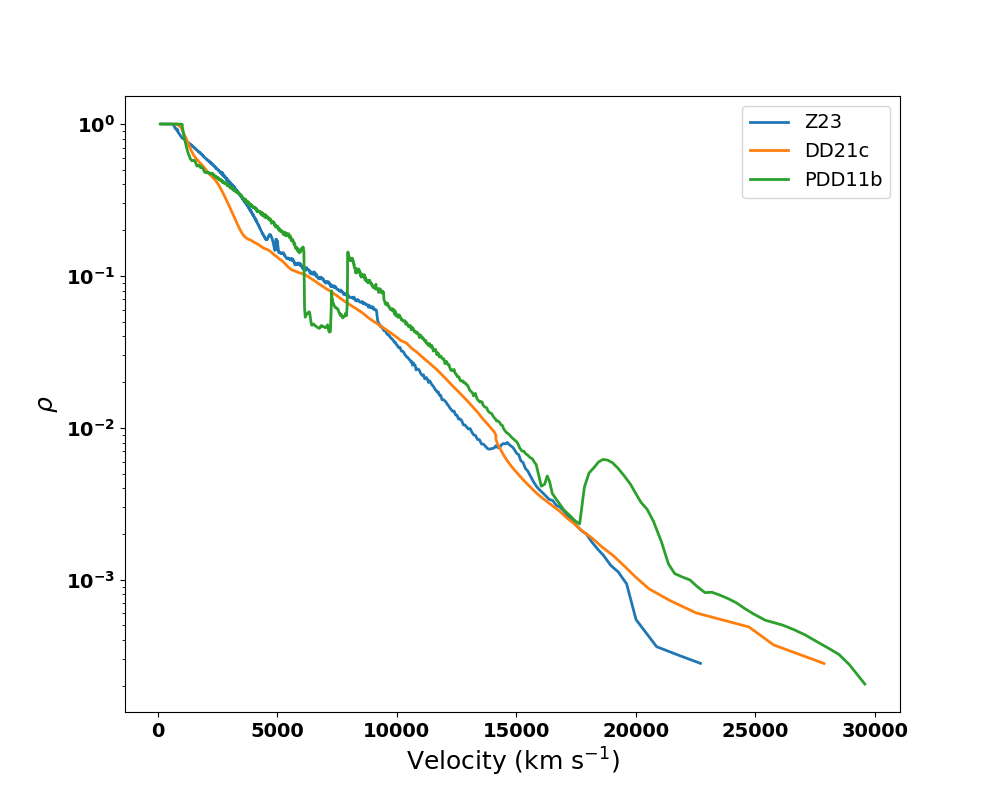}}
\subcaptionbox{Z23}{\includegraphics[width=0.45\linewidth]{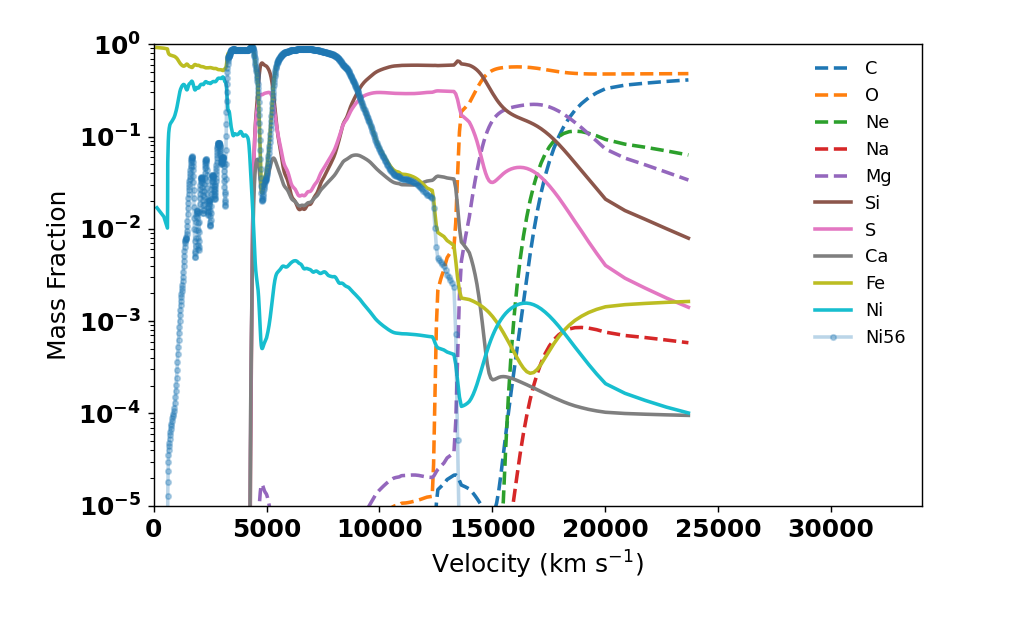}}
\subcaptionbox{DD21c}{\includegraphics[width=0.45\linewidth]{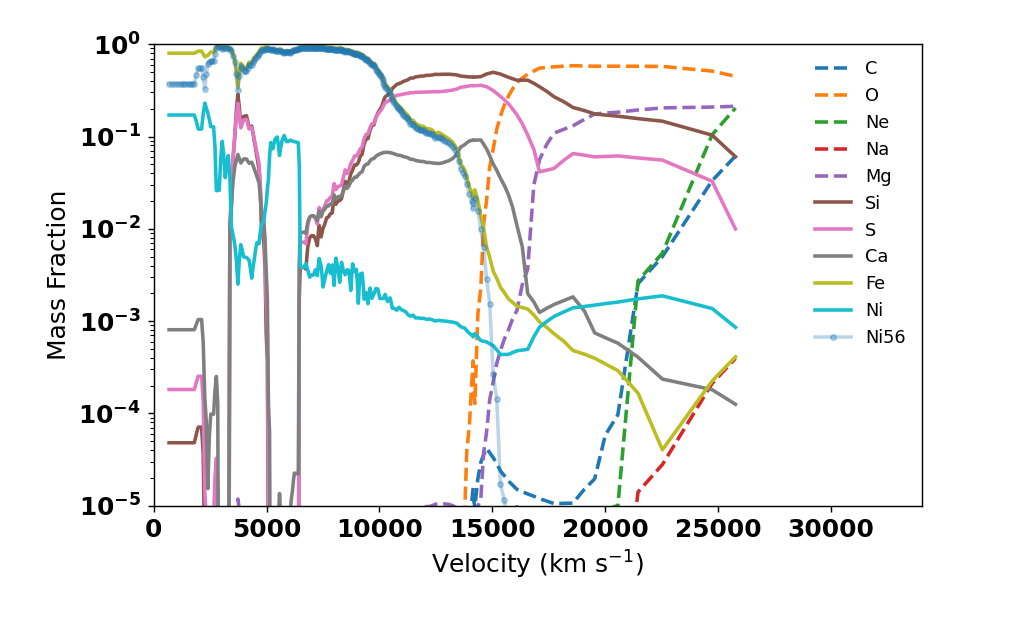}}
\subcaptionbox{PDD11b}{\includegraphics[width=0.45\linewidth]{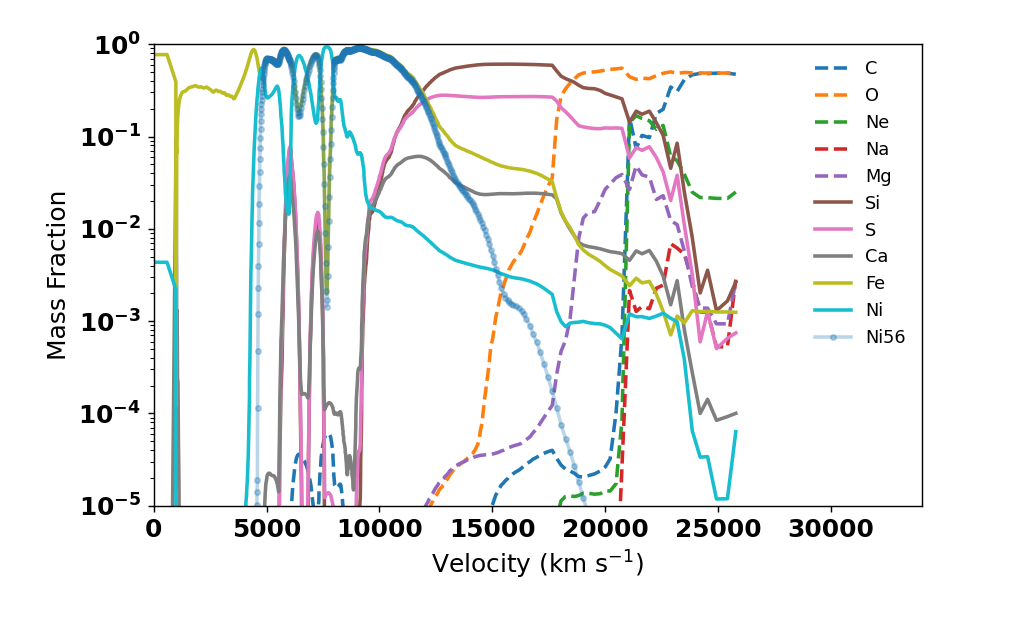}}
\caption{The density and composition profiles  of the two delayed
  detonation models 
  DD21c, Z23, and the pulsational delayed-detonation model PDD\_11b 
  (used to model SN~1984A, SN~2011fe,
  and SN~2001ay, respectively)
  are compared. All elements except $\nnni$ are displayed at
  $t=\infty$, while $\nnni$ is given at $t=0$.}
\label{fig:dd21c-z23}
\end{figure*}

\autoref{fig:dd21c-z23}(a) shows the density profiles of our
previously calculated models for SN~2011fe (Z23), SN~1984A (DD21c),
and SN~2001ay (PDD\_11b). While the pulsational delayed-detonation
model clearly displays a shell as expected, the obvious differences
between the BL models and the CN model are: higher densities at higher
velocities for the BL models; significant carbon depletion at high
velocities for the BL models; and a more extended $\nnni$ distribution
for the BL models although the extremely extended $\nnni$ distribution
for PDD\_11b is a particular feature that was needed to reproduce the
observed light curve shape of SN~2001ay \citep{Baron:2012:01ay}. It is
interesting to note that most BL and HV SNe Ia have no observed
\ion{C}{II} absorption at early times \cite[see, for
  example,][]{Parrent:2012:11fe}, with the notable exception of
SN~2019ein.

In general, high expansion velocities can be formed in two ways.
First, by the redistribution of the kinetic energy by means of a shell
interaction, namely the ejecta run into a dense circumstellar medium~(CSM)
shell produced during a large amplitude pulsational phase;
the characteristics of this CSM shell interaction are slow-rising-red
lightcurves and early blue line wings with a almost linear relation
between the blue edge of the velocities of quasi-statistical
equilibrium (QSE) elements and the mass of the shell
\citep{Khokhlov_etal_1992,Hoeflich_Khokhlov_1996,Quimby_etal_2006}.
The separation in the production of a high velocity shell between
classical delayed-detonation and pulsational delayed-detonation is
not a bifurcation, but rather a continuum with low mass, low amplitude
pulsations appearing to be most common \citep{Hoeflich_etal_2017}.

 The second mechanism is related to an increase of nuclear energy 
  as a result of the C/O ratio produced during the central He-burning under
  He-depleted conditions as suggested for SN2001ay
  \citep{Baron:2012:01ay}. Within the framework of delayed-detonation
  models, the results are 
  high velocity wings in Si as commonly observed in SNe~Ia, but also overall
  broad lines in features produced in QSE. The size of the
  He-burning zone 
  depends on the main-sequence mass of the progenitor and ranges between
  $0.2 - 1.0$~\msol  for a  MS mass of $1 - 8$~\msol
  \citep{Dominguez_etal_2001}. 
  Thus, we may expect a continuum of expansion velocities for BL SNe.

For SN~2019ein, the outermost layers show C/O rich mixtures, but also a
wing of Si extending up to about 24,000~\kmps
\citep{Patra:Yang:2019ein:2022} implying a 
upper mass limit 
of a shell to about  $5 \times 10^{-2}$~\msol
\citep{Yang_etal2020_ApJ_902_46}. Thus, this suggests that
  SN~2019ein falls in the continuum of low mass, low amplitude
  pulsations \citep{Hoeflich_etal_2017}.

\subsection{The BL Sample}
\label{sec:sample}

\begin{figure}
\centering
\includegraphics[scale=0.3]{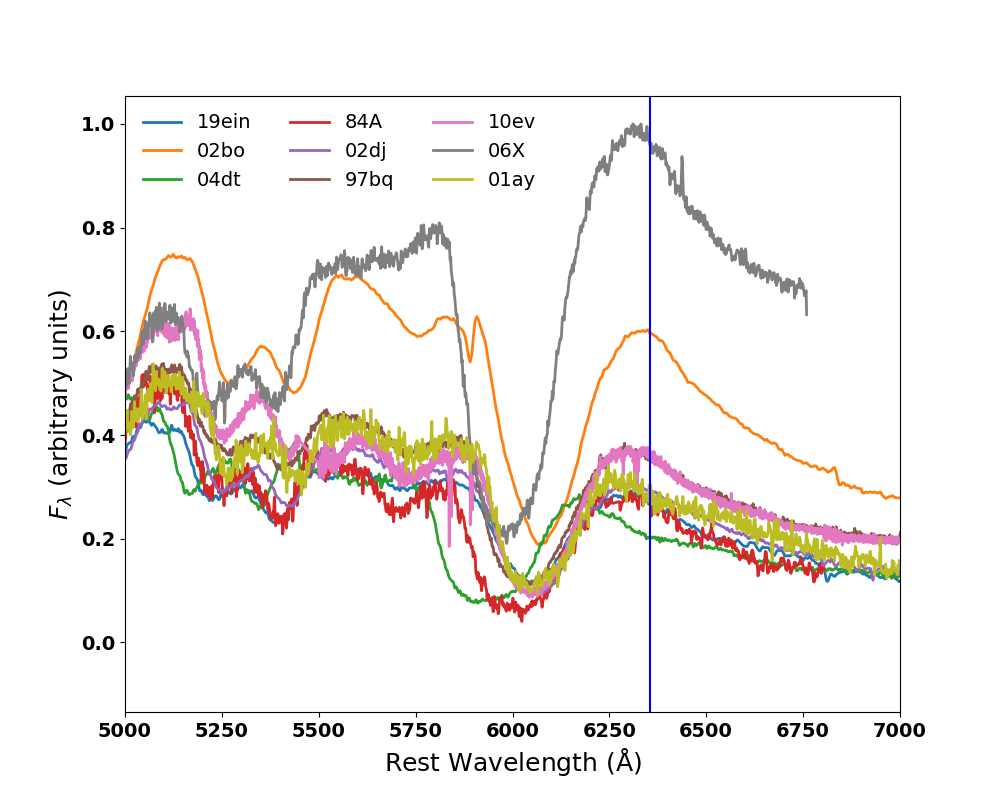}
\caption{The  spectra of BL SNe~Ia observed at days -2 -- -6
  with respect to $B$-maximum. The dates are (-5, -6, -2, -4, 
  -5, -2, -6, -4, and -4)
  for (84A, 97bq, 01ay, 02bo, 02dj, 04dt, 06X, 10ev, and 19ein),
  respectively. See \autoref{sec:sample}
   for the original sources of these 
  spectra.
      With the exception of SN 2002bo, all of the BL/High-velocity SNe show
  that the P-Cygni emission
  peak of \ion{Si}{II} $\lambda 6355$ is blueshifted in the epochs -2
  -- -6, thus this could serve as a simple classification criterion.}
\label{fig:02bocompare}
\end{figure}

We collected
spectra of BL SNe that were obtained prior to maximum light in the
$B$-band, including: SNe 1984A \citep{Branch1987,Barbon1989},
1997bq \citep{Blondin_etal_2012}, 2001ay \citep{Krisciunas:2011},
2002bo \citep{Benetti2004}, 2002dj \citep{Pignata2008}, 2004dt
\citep{Altavilla:2007}, 2006X \citep{Wang:2008:06X}, 2010ev
\citep{Gutierrez:2016}, and 2019ein 
\citep{Kawabata:Maeda:2019ein:2020,Pellegrino:2020:19ein}.
\autoref{fig:02bocompare} shows the \ion{S}{II}  and \ion{Si}{II}
regions of these supernova spectra.
All of the BL/High-velocity SNe that we have examined show
  that the 
  P-Cygni emission
  peak of \ion{Si}{II} $\lambda 6355$ is blueshifted in the epochs -6
  -- -2 days, thus this could serve as a simple classification
  criterion (\autoref{fig:02bocompare}). The validity of this indicator as a classification
    criterion will be the subject of future work. If we use the
  preferred redshift suggested by NED\footnote{\url{https://ned.ipac.caltech.edu/}} for the 
  SN~2002bo host NGC~3190, $z=0.00437$, \citep[$v$ heliocentric = 1310
  \kmps][]{2016MNRAS.463.1692L}, the peak of SN~2002bo is not
  blue-shifted. This seems unlikely that  the class defining -4 day
  spectrum of SN~2002bo would be 
  the one exception, so therefore we have adopted the higher value $v$
  heliocentric = 1698  \kmps \citep{2000ApJ...529..786M}. NGC~3190 is
  a very nearby, nearly edge-on spiral, so a peculiar velocity of
  300~\kmps in the galaxy for the supernova would not be surprising.

\begin{figure}
\centering
\includegraphics[width=0.49\textwidth]{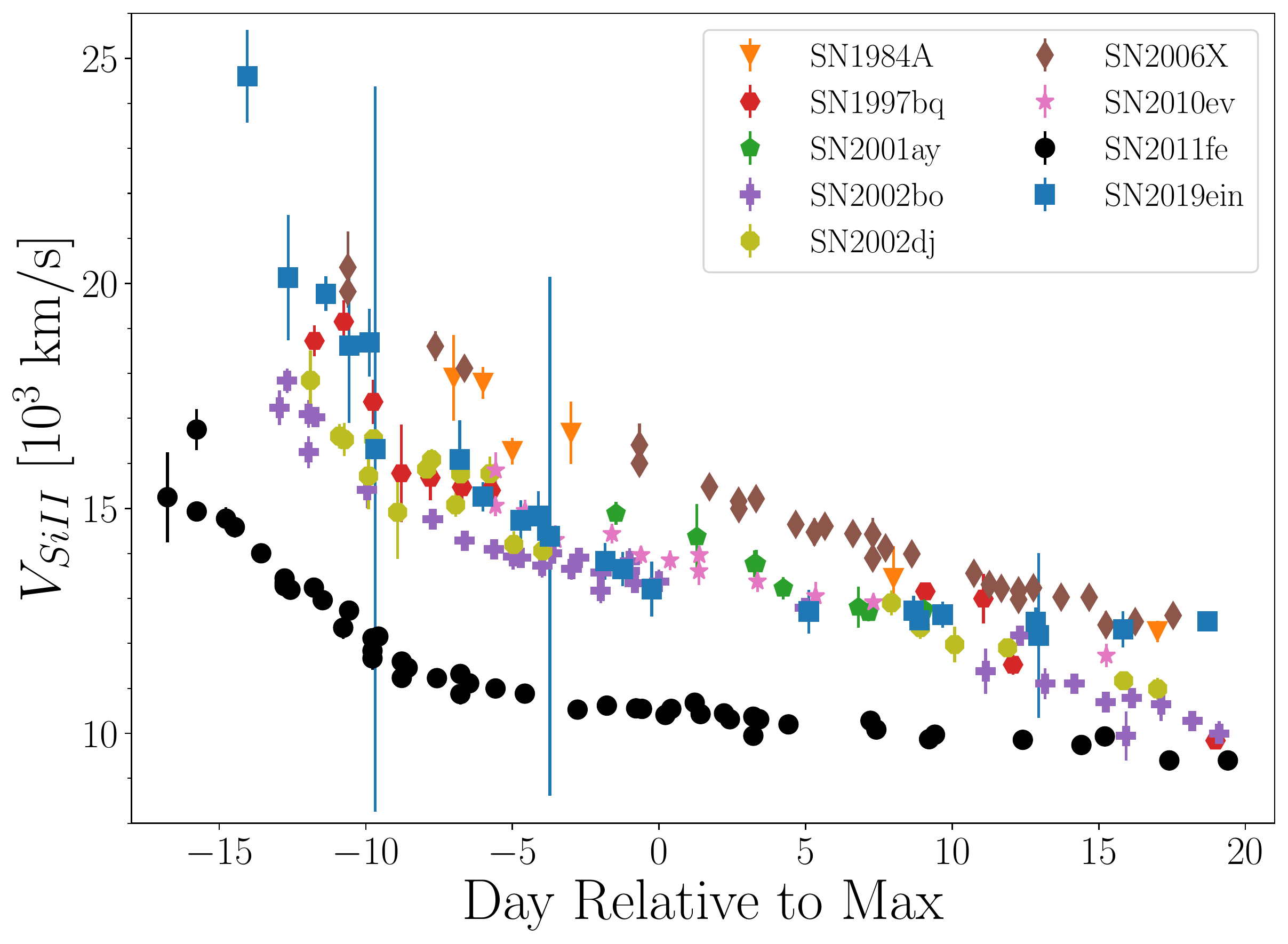}
\caption{The  measured velocity of the the \ion{Si}{II}
  $\lambda 6355$ absorption trough minimum  for the set of BL-SNe~Ia, including
  SNe~1984A \citep{Branch1987,Barbon1989}, 1997bq \citep{Blondin_etal_2012}, 
  2001ay \citep{Krisciunas:2011}, 2002bo \citep{Benetti2004}, 
  2002dj \citep{Pignata2008}, 2006X \citep{Wang:2008:06X}, 
  2010ev \citep{Gutierrez:2016}, and 
  2019ein \citep{Kawabata:Maeda:2019ein:2020,Pellegrino:2020:19ein}. 
  Measurements of SN~2011fe \citep{Parrent:2012:11fe,Pereira:2013:11fe} are
  also shown for comparison.}
\label{fig:02bovels}
\end{figure}

\autoref{fig:02bovels} shows the velocities of the Si~II absorption
trough minimum for a set of BL SNe~Ia.  This
  figure shows behavior very similar to that found by
  \citet{Wang:2009} where there is a continuum of velocities with SN
  2002bo being one of the lower velocity SN and SN~2006X being one of
  the higher velocity SN. This is consistent with the discussion above
  that suggests in the framework of delayed detonation, a continuum of
  line velocities is expected. Note that even at 10--15 days past maximum
  light, the Si velocities have not fallen to the values of SN~2011fe,
  clearly indicating that even though the nucleosynthetic products of
  BL SNe~Ia are nearly identical to that of CN, BL SNe~Ia are a
  distinct sub-class. 

\begin{figure}
\centering
\subcaptionbox{}{\includegraphics[scale=0.35]{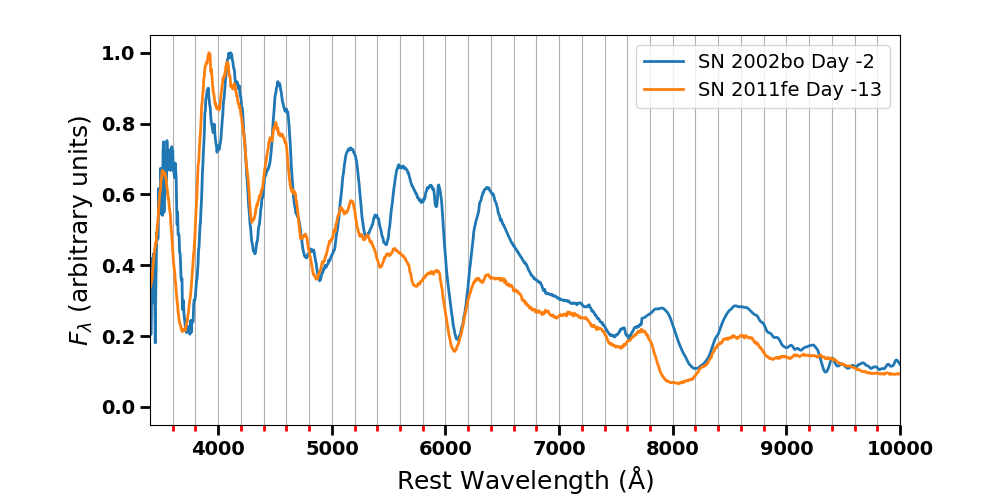}}
\subcaptionbox{}{\includegraphics[scale=0.35]{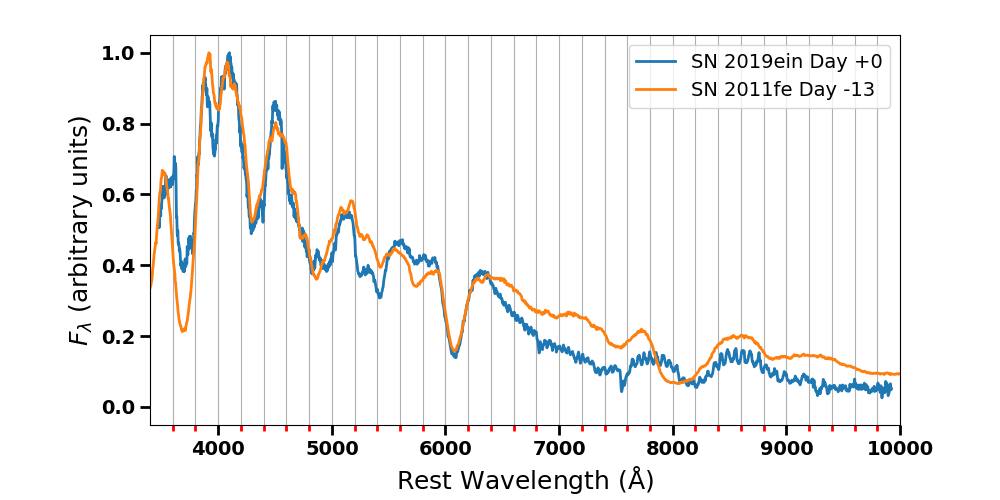}}
\subcaptionbox{}{\includegraphics[scale=0.35]{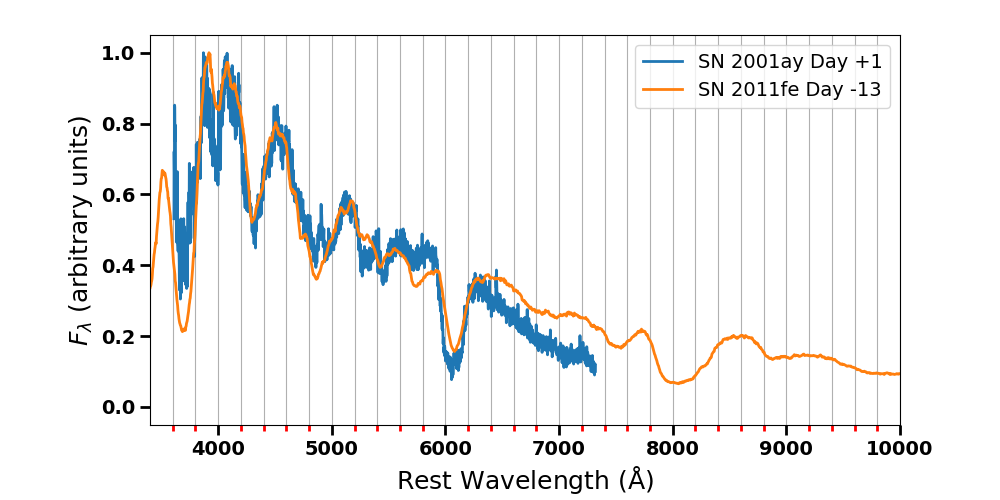}}
\caption{Panel (a) compares SN~2002bo at day -2 to SN~2011fe at day
-13. Panel (b) compares SN~2019ein at day +0 to SN~2011fe at day
-13. Panel (c) compares SN~2001ay at day +1 to SN~2011fe at day
-14. A broad variety of
BL supernovae have observed spectra near
maximum light are well matched by early-time spectra of the CN
SN~2011fe, suggesting that the high observed velocities are truly due
to variations in the outermost part of the ejecta of BL SNe.}
\label{fig:date-offset}
\end{figure}

\autoref{fig:date-offset} shows that for a broad variety of
BL/High-Velocity/02bo-like supernovae that the observed spectra near
maximum light are well matched by early-time spectra of the CN
SN~2011fe. Interestingly, the poorest match is for the class defining
SN~2002bo, where the Ca IR triplet of SN~2011fe extends to higher
velocity than that of SN~2002bo and the strength of the \ion{Si}{II}
lines is not as well-matched as it is for SN~2019ein and
SN~2001ay. Unfortunately, the maximum light spectrum of SN~2001ay does
not extend to the Ca IR triplet, but the fit in the blue is
remarkable.

\citet{Xi:2022:19ein} found that the high velocities of  SN~2019ein
are due to asymmetries, however 
SN~2019ein shows no evidence for polarization
\citep{Patra:Yang:2019ein:2022} indicating that there are no very
strong asymmetries. In the delayed detonation framework, CN SNe~Ia
appear to arise from off-center detonations \citep{Derkacy_etal_2023_01aefx}. In looking for the progenitor/explosion scenario one
should note that BL occur preferentially in luminous galaxies
\citep{Folatelli_etal_2013,Morrell_etal_2023}.  

\subsection{Summary}
\label{sec:summ}

Our direct analysis shows that blue-shifted P-Cygni profiles can be
produced in the SYNOW framework by high velocity photospheric features
coupled with one or more detached component. This result is consistent
with detailed models of SN~1984A \citep{Lentz:2001} and SN~2001ay
\citep{Baron:2012:01ay} using a delayed-detonation and pulsational
delayed-detonation model respectively, showing
that shell like structures occur in the near-$M_\text{Ch}$ framework. 
The modeling shows that the outer compositions and ionization stages
of the core-normal SN 2011fe and the  broad-line SN
2019ein are extremely similar, except that the features of the BL SN
are blue-shifted with respect to those of the CN SN. This illustrates
the fact that the progenitor compositions must be very similar and
the burning leads to a so-called stellar amnesia
\citep{Hoeflich:2006:NP-Review}. Our models indicate the outer parts
of SN 2019ein likely form in a higher-density environment, something like
a shell, but the effects of the
shell is mostly dissipated by the time of maximum
light. Future work should focus on a comparative study of BL SNe~Ia
and CN in order to probe the progenitor variations between
these two Branch groups.

\clearpage

\section*{Acknowledgments}

We thank Ariel Goobar and Peter Nugent for helpful discussions.
E.B. and J.D. are supported in part by NASA grant
80NSSC20K0538.
Some of the  calculations presented here were performed at the
H\"ochstleistungs Rechenzentrum Nord (HLRN), 
at the National Energy Research Supercomputer Center (NERSC), which is
supported by the Office of Science of the U.S.  Department of Energy under
Contract No. DE-AC03-76SF00098 and at
the OU Supercomputing Center for Education \& 
Research (OSCER) at the University of Oklahoma (OU).
We thank all these institutions for a generous
allocation of computer time.

\section*{Data Availability}

The observed spectra are available from
\href{https://www.wis-tns.org/}{TNS},
\href{https://www.wiserep.org/}{WISeREP} \citep{Yaron_GalYam_2012} and the SYNOW fits are
available from the authors upon request.

\bibliographystyle{mnras}
\bibliography{zach19ein_bib}

\bsp	\label{lastpage}
\end{document}